\documentclass[aps,prl,preprint,superscriptaddress]{revtex4-1}

\usepackage{graphicx}
\usepackage{amssymb, amsmath}
\usepackage[usenames]{color}
\usepackage{url}
\def\k{\mathbf{k}}
\def\r{\mathbf{r}}

\begin{document}
\title{Topological nodal-line semimetals in alkaline-earth stannides, germanides and silicides}

\author{Huaqing Huang}
\affiliation{Department of Physics and State Key Laboratory of Low-Dimensional Quantum Physics, Tsinghua University, Beijing 100084, China}
\affiliation{Collaborative Innovation Center of Quantum Matter, Tsinghua University, Beijing 100084, China}

\author{Jianpeng Liu}
\affiliation{Kavli Institute for Theoretical Physics, University of California, Santa Barbara, California 93106, USA}
\affiliation{Department of Physics and Astronomy, Rutgers University, Piscataway, New Jersey  08854-8019, USA}

\author{David Vanderbilt}
\affiliation{Department of Physics and Astronomy, Rutgers University, Piscataway, New Jersey  08854-8019, USA}

\author{Wenhui Duan\footnote{dwh@phys.tsinghua.edu.cn}}
\affiliation{Department of Physics and State Key Laboratory of Low-Dimensional Quantum Physics, Tsinghua University, Beijing 100084, China}
\affiliation{Collaborative Innovation Center of Quantum Matter, Tsinghua University, Beijing 100084, China}
\affiliation{Institute for Advanced Study, Tsinghua University, Beijing 100084, China}
\date{\today}

\begin{abstract}
Based on first-principles calculations and an effective Hamiltonian analysis,
we systematically investigate
the electronic and topological properties of alkaline-earth compounds $AX_2$
($A$\,=\,Ca, Sr, Ba; $X$\,=\,Si, Ge, Sn). Taking BaSn$_2$ as an example, we find that
when spin-orbit coupling is ignored,
these materials are three-dimensional topological nodal-line semimetals
characterized by a snake-like nodal loop in three-dimensional momentum
space.  Drumhead-like surface states
emerge either inside or outside the loop circle on the (001) surface depending
on surface termination, while complicated double-drumhead-like surface
states appear on the (010) surface. When spin-orbit coupling is included, the
nodal line is gapped and the system becomes a topological insulator
with $\mathbb{Z}_2$ topological invariants (1;001). Since spin-orbit coupling effects are
weak in light elements, the nodal-line semimetal phase
is expected to be achievable in some alkaline-earth germanides and silicides.
\end{abstract}

\maketitle

\paragraph*{Introduction.}
The discoveries of  topological insulators \cite{RevModPhys.82.3045,
*RevModPhys.83.1057, PhysRevB.90.195105} and Chern insulators
\cite{Haldane, chang2013experimental, PhysRevB.90.165143, semiDirac}
have attracted broad interest to topological aspects of band theory.
These topologically nontrivial materials are bulk insulators with novel
gapless edge or surface states protected by bulk band topology. Recently
it has been shown that certain types of bulk semimetallic systems may also
possess non-trivial topological properties such as topologically protected
gapless surface states. These topological semimetallic
states are characterized by band touching (BT) points or lines between
valence and conduction bands in three-dimensional (3D)
momentum space. Up to now, three types of topologically nontrivial
semimetals have been proposed: Dirac \cite{Na3Bi,*Cd3As2}, Weyl
\cite{TaAsPRX,*TaAsHLin,*lv2015discovery,*xu2015experimental}
and nodal-line semimetals \cite{TNS,BernalGraphite,*BernalGraphite2,
heikkila2011dimensional,*heikkila2015nexus,MTC,Cu3PdN,
*Cu3NNi,PbTaSe2,*TlTaSe2,xie2015new,chan2015topological}.
Weyl and Dirac semimetals exhibit two-fold
and four-fold degenerate BT points near the Fermi level respectively,
and their low-energy bulk excitations are linearly dispersing Weyl or
Dirac fermions. Unlike Weyl and Dirac semimetals with isolated bulk BT
points, topological nodal-line semimetals possess BTs along one-dimensional
(1D) loops or lines in 3D momentum space.

So far, significant progress has been made in identifying topological
semimetals in realistic materials. For example, the Dirac-semimetal
state in Na$_3$Bi and Cd$_3$As$_2$ \cite{Na3Bi,*Cd3As2} and Weyl-semimetal
state in the TaAs family of compounds
\cite{TaAsPRX,*TaAsHLin,*lv2015discovery,*xu2015experimental} have been
predicted theoretically and then verified by experiments. A few
candidate materials for topological nodal-line semimetals have also
been proposed recently. Depending on the degeneracy of the states along the
loop, these materials can be classified into two groups.
One is known as the Dirac-type nodal-line semimetal, in which two doubly
degenerate bands cross each other to form a four-fold degenerate nodal loop.
This kind of nodal-line semimetal can typically be realized in materials
with both inversion and time reversal symmetries when spin-orbit coupling (SOC)
is neglected and band inversion happens around one or more high-symmetry
points in Brillouin zone (BZ). The other one lacks either inversion or
time-reversal symmetry, so that the otherwise four-fold degenerate nodal loops
are split into two doubly degenerate nodal loops,
called ``Weyl nodal lines'' in the literature \cite{chan2015topological}.
These nodal lines are usually protected by additional crystalline
symmetries such as mirror reflection so as to be stable against
perturbations, including SOC.

In the present work, based on first-principles calculations and a model
Hamiltonian analysis, we find a topological nodal-line semimetal phase
in the alkaline-earth compound $AX_2$ ($A$\,=\,Ca, Sr, Ba; $X$\,=\,Si, Ge, Sn) when SOC is
absent. Unlike the existing nodal-line semimetal materials, this
system exhibits a snake-like nodal loop in 3D momentum space. More
interestingly, dispersive gapless surface states exist on the (001) surface,
while some special double-drumhead-like surface states occur on the (010)
surface. Including SOC in the first-principles calculation leads to a
transition from a nodal-line semimetal to a strong topological insulator
with (1;001) $\mathbb{Z}_2$ indices.
Since the SOC effect is negligible in silicides and remains small in
germanides, the nodal-line semimetal phase is expected to be
realized in these compounds.

\begin{figure}
\includegraphics[width =0.9\columnwidth]{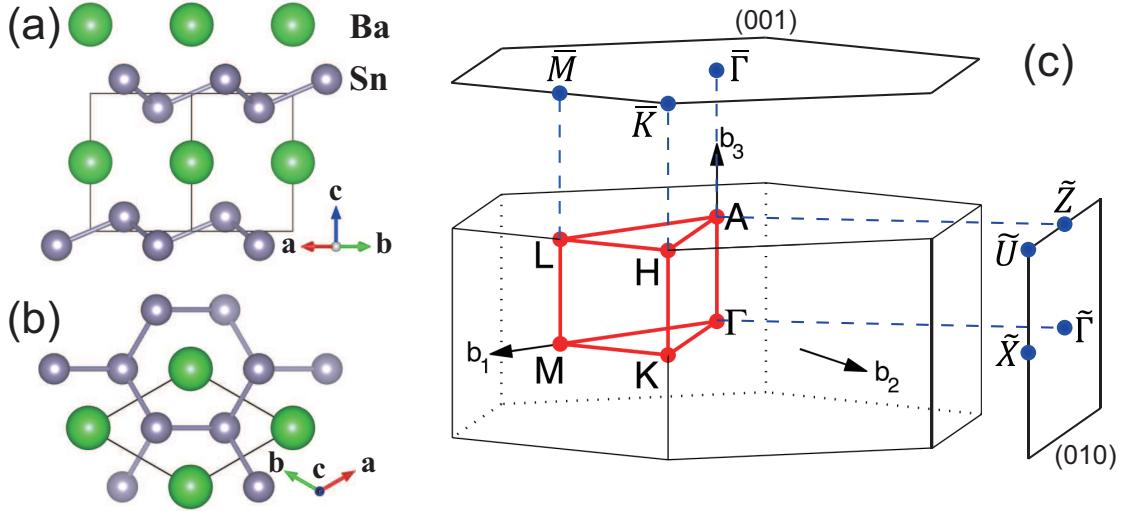}%
\caption{\label{fig1_stru} (Color online) (a)-(b) Crystal structure
of BaSn$_2$ with $P\bar{3}m1$ symmetry. (c) Brillouin zone of the bulk
and the projected surface Brillouin zones of the (001) and (010) surfaces.}
\end{figure}

\paragraph*{Crystal structure and methodology.\label{}}
Since the compounds in the $AX_2$ family have similar centrosymmetric crystal
structures and electronic structures, we take BaSn$_2$, which has been
synthesized successfully \cite{kim2008crystal, ropp2012encyclopedia},
as an example hereafter. BaSn$_2$ crystallizes in the trigonal $P\bar{3}m1$
structure (space group No.~164) as shown in Fig.~\ref{fig1_stru}.
In this structure, the Sn atoms form a buckled honeycomb lattice,
while each hexagonal Ba layer intercalates between two neighboring Sn layers.

In order to understand the electronic structure of the $AX_2$ compounds,
we perform first-principles calculations within the framework of
density-functional theory as implemented in the Vienna {\it ab initio}
simulation package \cite{VASP} with the projector augmented-wave method \cite{PAW}.
The Perdew-Burke-Ernzerhof exchange-correlation funcional in the
generalized gradient approximation is adopted \cite{PBE}.
The kinetic
energy cutoff is fixed to 520\,eV and an 8$\times$8$\times8$
$\Gamma$-centered $\mathbf{k}$ mesh is used for the BZ sampling.
The electronic structure is calculated both with and without SOC.
We also generate \textit{ab initio} tight-binding Hamiltonians in the
basis of projected atomic-like Wannier functions without applying a
maximal-localization procedure \cite{wannier1, *wannier2, wannier90}.
In the Wannier representation, we calculate the surface states
using the iterative Green's function method \cite{lopez,*lopez2} based on the bulk tight-binding
model neglecting possible surface reconstructions and charge rearrangements. The topological
feature of the surface states are expected to be captured by
such simplified Wannier-based approach
\footnote{\label{fn}See Supplementary Materials,
http://link.aps.org/supplemental/xxx,
for more details about the computational methods, the surface states, the
effect of SOC, and the results of calculations of the electronic
structures of other $AX_2$ compounds.}.

\begin{figure}
\includegraphics[width =0.9\columnwidth]{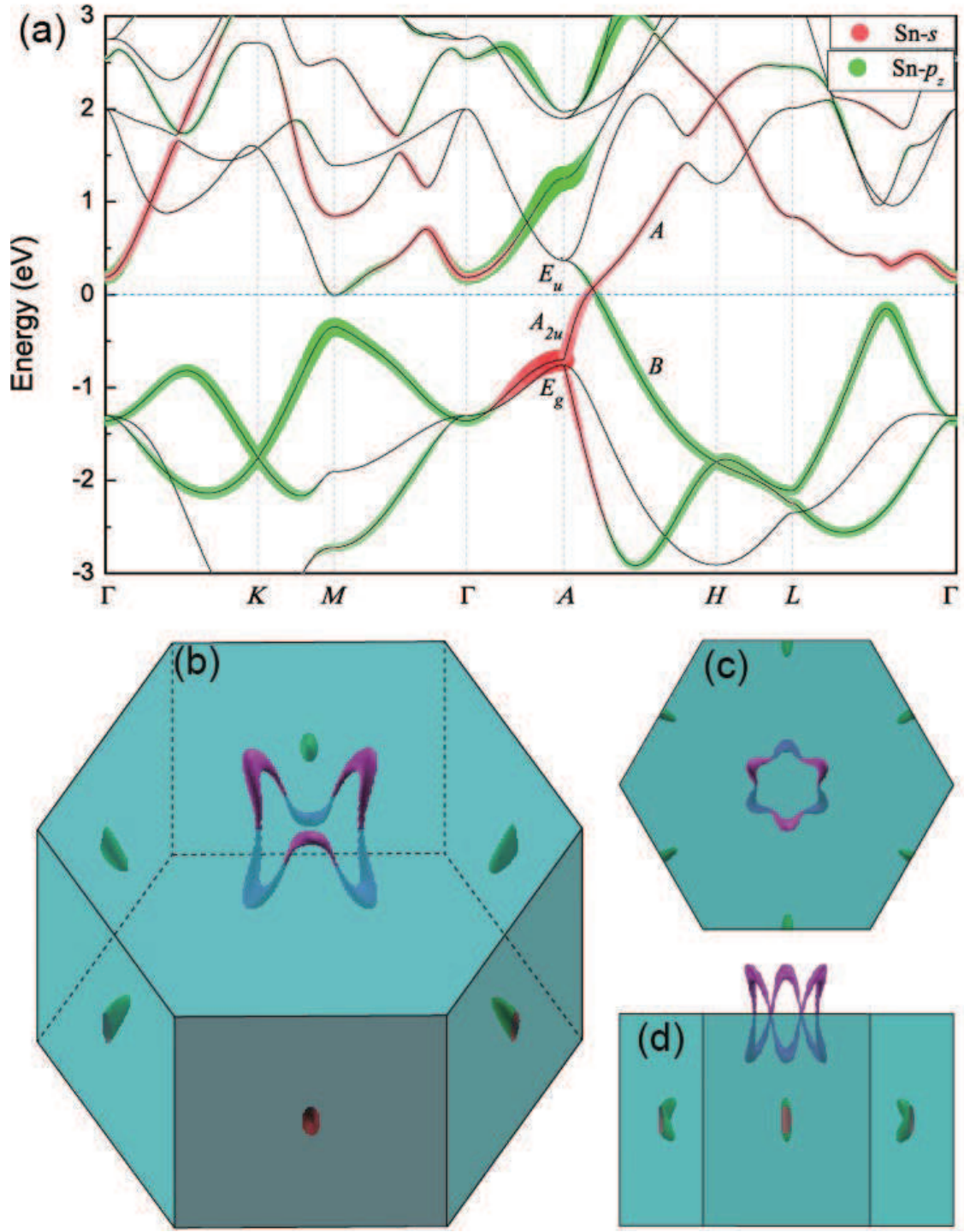}%
\caption{\label{fig2_fermi} (Color online) (a) Calculated band structure
without SOC. The irreducible representation of selected bands at $A$ and along $A$-$H$
line are indicated. Red and green dots in band structures indicate
the projection to the Sn $s$ and $p_z$ orbital, respectively.
(b) Corresponding 3D iso-energy surface at
$E=E_F+25$\,meV. (c) Top view and (d) side view of the isoenergy surface
from the (001) and (010) directions of the Brillouin zone, respectively.}
\end{figure}

\paragraph*{Electronic structures.}
We first study the electronic structure of BaSn$_2$ in the absence of SOC.
The band structure along the high-symmetry path marked by red
lines in Fig.~\ref{fig1_stru}(c) is shown in Fig.~\ref{fig2_fermi}(a).
It can be seen that there is a BT close to the Fermi
level along the $A$-$H$ line.
Since the lowest conduction band and the
highest valence band along this line have opposite parities with respect
to the $C_2$ operation in the little group of $A$-$H$, they can touch
each other without opening a gap. Further orbital-character analysis
reveals that the states around the Fermi level are dominated by Sn~$s$
and Sn~$p_z$ orbitals, as shown in Fig.~\ref{fig2_fermi}(a).
Moreover, there is a
clear signature of band inversion at the $A$ point, implying possible
non-trivial band topology. It should be noted that the band inversion
at the $A$ point is not due to SOC as it is excluded in this calculation.

Although the linear dispersion around the BT resembles the character
of a Dirac semimetal, the system is actually a nodal-line semimetal.
Namely, the BT persists along a closed loop in the 3D BZ forming a snake-like
nodal loop around the $A$ point. If the loop were constant in energy,
the Fermi surface would shrink exactly to the nodal loop. While the energy of the
BT varies slightly along the loop, its shape is still easily
inferred by looking at an isoenergy countour at an energy slightly
above that of the loop.
Such a contour, plotted at 25\,meV above the Fermi energy $E_F$,
is shown in
Fig.~\ref{fig2_fermi}(b)-\ref{fig2_fermi}(d). Given that the BZ
is periodic, the calculated isoenergy surface forms a a closed
snake-like pipe enclosing the nodal line which in turn winds around
the $A$ point. Thus, there is a single nodal loop in the BZ.

Further fine band structure calculations within the 1/6
BZ using Wannier interpolation indicate that the nodal loop
has its maximum energy of 0.117\,eV when passing through the
$\Gamma$-$M$-$L$-$A$ plane at
$(0.136, 0.0, 0.707)$ (in units of $(2\pi/a, 2\pi/a, 2\pi/c)$),
and its minimum of 0.064\,eV when passing through the $A$-$H$ line
at (0.064, 0.064, 0.5).  The energy then increases back to 0.117\,eV
as the loop snakes
to $(0.0, 0.136, 0.293)$ on the adjacent $k_y$-$k_z$ plane, and
the remaining 5/6 of the loop is related to this segment by
symmetries.
BaSn$_2$ is unusual in the sense that it has a single nodal loop, whereas
other nodal-loop systems such as Cu$_3$NPd(Zn) \cite{Cu3PdN,*Cu3NNi}
and Pb(Tl)TaSe$_2$ \cite{PbTaSe2,*TlTaSe2}
typically have multiple circles or ellipses.
Moreover, despite three small electron pockets around the $M$ point,
the Fermi surface is mostly
contributed by the single nodal loop, which makes it easier to study
any intriguing properties related to the presence of the nodal loop.

\paragraph*{Symmetry analysis.}
The nodal loop in BaSn$_2$ is topologically protected due to the
coexistence of time-reversal ($\mathcal{T}$) and spatial inversion
($\mathcal{P}$) symmetries \cite{MTC,martin2004electronic}. When SOC
is absent, the system can be considered as a spinless system
for which $\mathcal{T}$ is simply a complex conjugation operator.
Therefore
one can adopt a gauge {for the Bloch functions} such that $u_{n\k}^{*}(\r)\!=\!u_{n-\k}(\r)$.
On the other hand, inversion symmetry connects  $u_{n\k}(-\r)$ to $u_{n-\k}(\r)$, and
we are allowed to let $u_{n\k}(-\r)\!=\!u_{n-\k}(\r)$.
Combining the above two equations, one obtains $u_{n\k}^*(\r)\!=\!u_{n\k}(-\r)$.
Then it is straightforward to show that the corresponding effective
Hamiltonian $H(\k)$ has to be real-valued, i.e.,
$H_{mn}(\k)\!=\!H_{nm}(\k)$.

A BT problem at an arbitrary $\k$ point can be minimally described
by a two-band effective Hamiltonian, which can be expressed in terms
of the identity matrix and the three Pauli matrices. According to the
above argument, the two-band Hamiltonian for a system respecting both
$\mathcal{T}$ and $\mathcal{P}$ symmetries can be chosen as real valued,
so the codimension of such a BT problem is 2, one less than the number
of independent variables (i.e., $k_x$, $k_y$ and $k_z$). Hence a nodal
loop is stable in the presence of coexisting $\mathcal{T}$ and
$\mathcal{P}$ symmetries.

Some other symmetries of the system, such as $C_2$ rotation about the
$A$-$H$ line and $C_3$ rotation about the $k_z$ axis, are crucial in
determining the shape of the nodal loop.
Nevertheless, the nodal loop would be robust even in
the absence of these additional crystal symmetries; weakly breaking
them would just distort the loop.

\begin{figure}
\includegraphics[width = 0.9\columnwidth]{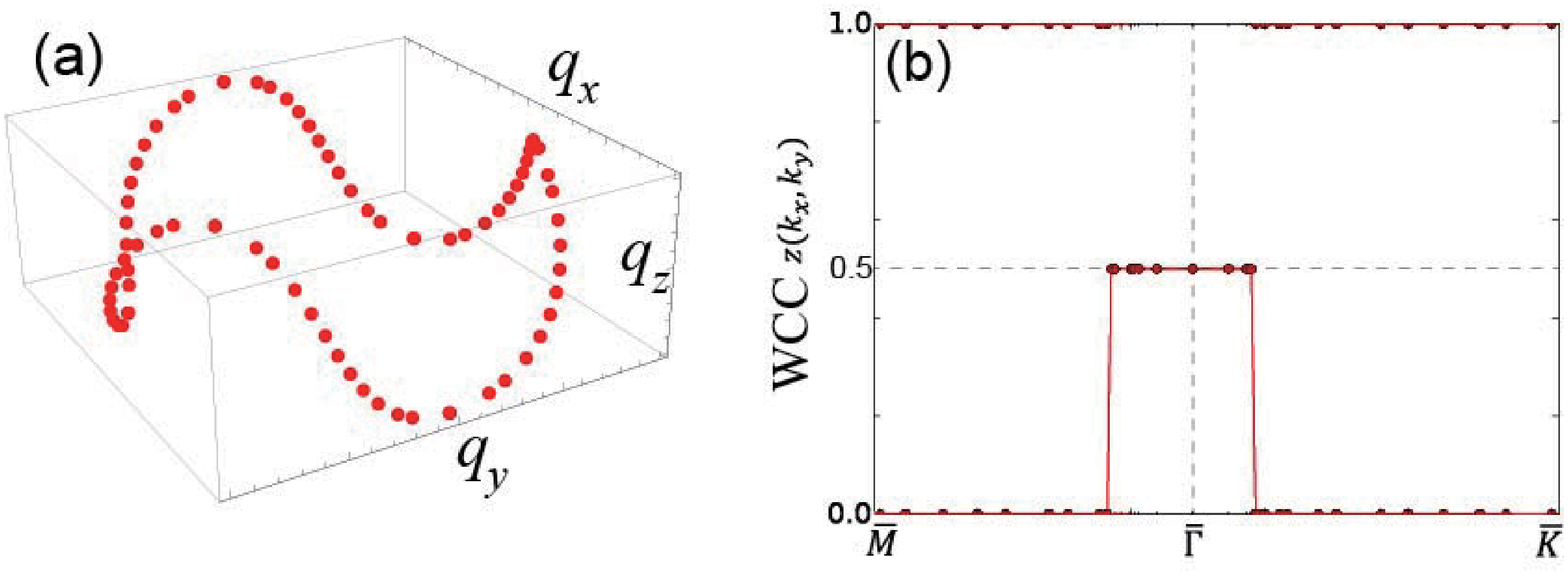}%
\caption{\label{fig3_kp} (Color online) Calculated band touching points
(nodal line) in 3D $\mathbf{k}$-space using the effective Hamiltonian
of Eqs.~(\ref{eq_def}) and (\ref{eq_ham}). (b) Evolution of Wannier charge
center $ z (k_x,k_y)$ for the occupied bands.}
\end{figure}

\paragraph*{Effective Hamiltonian.}
We further construct a minimal effective Hamiltonian that can
describe the nodal loop around the $A$ point.
We adopt a spinless two-band Hamiltonian of the form
\begin{equation}
h(\mathbf{q})=\epsilon(\mathbf{q})\sigma_0+\mathbf{d}(\mathbf{q})\cdot\vec{\sigma},
\label{eq_def}
\end{equation}
where the  $2\times2$ identity matrix $\sigma_0$ and Pauli matrices
$\vec{\sigma}=(\sigma_x,\sigma_y,\sigma_z)$ operate in the pseudospin
space of the two bands that cross near the $A$ point and
$\mathbf{q}=\mathbf{k}-{\bf k}_A$.
Since $\epsilon(\mathbf{q})\sigma_0$ represents an overall energy
shift, we neglect this term in the following analysis.

Let us start by considering the symmetries of the systems which impose
constraints on the allowed form of the coefficients $d_i(\mathbf{q})$
($i\!=\!x,y,z$). The little group at $A$ contains a threefold rotation
$C_3$ about $k_z$, a twofold rotation $C_2$ about the $A$-$H$ axis,
inversion $\mathcal{P}$, and time reversal $\mathcal{T}$.
Thus the Hamiltonian has to obey the symmetry constraints
$U\, h(\mathbf{q})\, U^{-1}=h(U^{-1}\mathbf{q})$, for $U=C_3, C_2,
\mathcal{P}$ and $\mathcal{T}$ \footnotemark[\value{footnote}].

First-principles calculations indicate that the low-energy states at
the $A$ point are mostly contributed by the Sn~$s$ and Sn~$p_z$ orbitals.
If we choose the two basis vectors as $|s\rangle$ and $|p_z\rangle$,
the symmetry operations of the little group take the form $C_3\!=\!\sigma_0$,
$C_2\!=\!\sigma_z$, $\mathcal{P}\!=\!\sigma_z$ and $\mathcal{T}\!=\!K\sigma_0$.
Then we obtain the symmetry-allowed expressions for $d_i(\mathbf{q})$:
\begin{equation}
\begin{split}
d_x &= 0,\\
d_y &= A q_z + B (q_x^2 q_y - q_x q_y^2) \\
    &\quad+ C (q_x^2 + q_y^2 - q_x q_y) q_z +D q_z^3,\\
d_z &= M+E(q_x^2 + q_y^2 - q_x q_y) + F q_z^2.
\end{split}\label{eq_ham}
\end{equation}
We see from Eq.~(\ref{eq_ham}) that the Hamiltonian is expressed only
in terms of two of the three Pauli matrices, which is consistent with
the previous codimension argument.

In order to achieve a band crossing on the $A$-$H$ line ($q_z\!=\!0$), $ME<0$
must be satisfied, which is nothing but the band-inversion condition.
The $A$ point lies within the band-inverted region, hence the two bands
tend to cross each other as $\k$ moves away from $A$. Based on the
effective Hamiltonian, we determine the BT points by solving the equations
$\mathbf{d}(\mathbf{q})\!=\!\mathbf{0}$. The calculated BT points form a
snake-like ring around the $A$ point as shown in Fig.~\ref{fig3_kp}(a),
which is consistent with the iso-energy-surface calculation as shown
in Fig.~\ref{fig2_fermi}(b).

\paragraph*{Topological invariant.}
To identify the nontrivial band topology in BaSn$_2$, we further study
the evolution of the 1D hybrid Wannier charge centers (WCCs) along a
high-symmetry path in the $k_x$-$k_y$ plane \cite{wannier1,*wannier2,
yuruiZ2,alexey1,*alexey2,maryam}.
The sum of the hybrid WCCs over the occupied subspace,
$z(k_x, k_y)$ (in unit of $c$), is equivalent
(up to a factor of $2\pi$) to the total Berry phase of the occupied Bloch
functions accumulated along the $k_z$ direction.
For a spinless system respecting both $\mathcal{T}$ and $\mathcal{P}$
symmetries, the Berry phase has to be either $0$ or $\pi$ for an
arbitrary loop in the BZ. Therefore, we expect that the total hybrid
WCC $z(k_x, k_y)$ would be quantized as either $0$ or $1/2$ at any
$(k_x, k_y)$,
and that $z(k_x, k_y)$ would jump by $1/2$ when passing through a
projected nodal point. This is exactly what happens in BaSn$_2$ without SOC.
As shown in Fig.~3(b), we find that $z(k_x, k_y)$ is 1/2 inside the
projected nodal loop centered at $\bar{\Gamma}$
in the projected 2D Brillouin zone and jumps to zero
outside, as is expected from the above discussions.

\begin{figure}
\includegraphics[width =0.9\columnwidth]{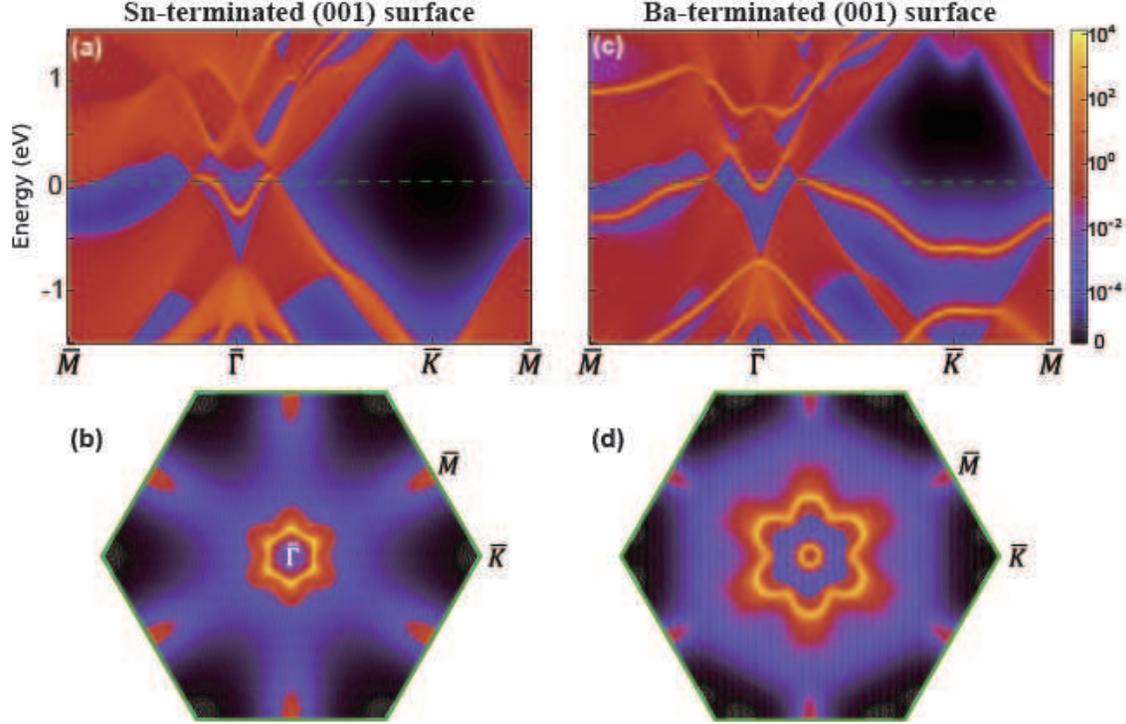}%
\caption{\label{fig4_surf} (Color online) Calculated (001) surface band
structure and Fermi surface without SOC for (a, b) Sn-terminated, and
(c, d) Ba-terminated, surfaces. The Fermi surfaces
shown in (b) and (d) are calculated with the chemical potential at
60\,meV [green dashed lines in (a) and(c)]. Note that the inner circle
in (d) originates from a concave conduction band state, not from the
drumhead-like surface state.}
\end{figure}

\begin{figure}
\includegraphics[width =0.9\columnwidth]{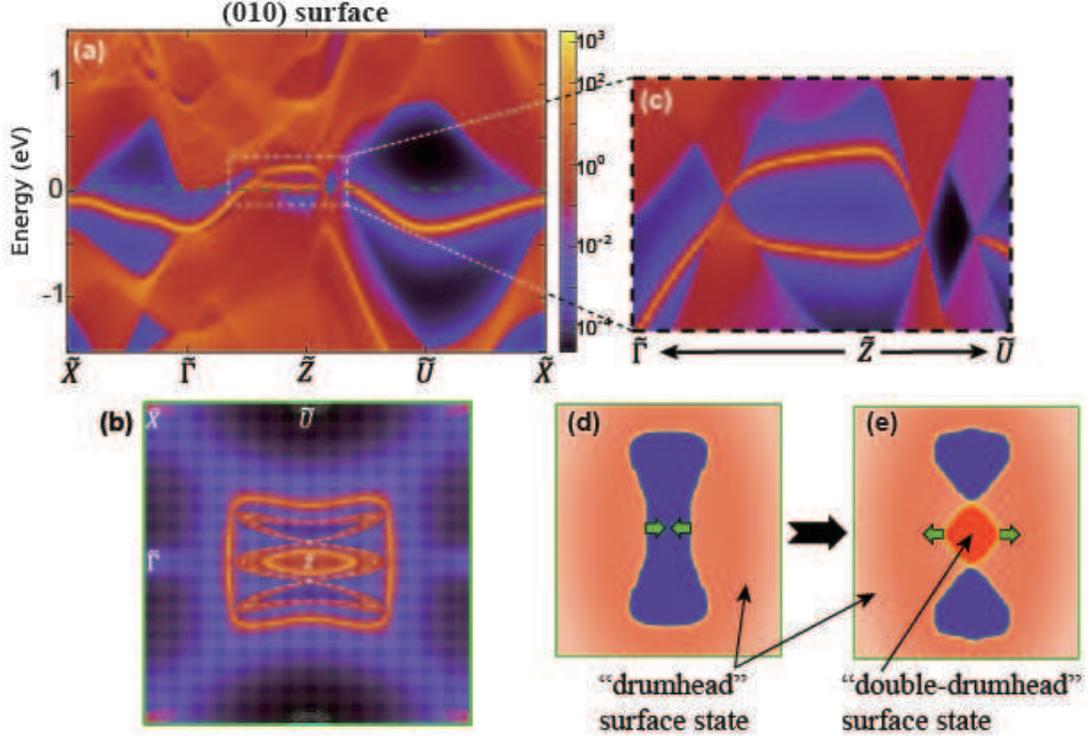}%
\caption{\label{fig5_surf} (Color online) (a) Calculated (010) surface band
structure and (b) Fermi surface without SOC. The
Fermi surfaces shown in (b) are calculated with the
chemical potential at 10\,meV [green dashed lines in (a)]. The white dashed
lines in (b) forming an ellipse and two crescents indicate the
projection of the bulk nodal loop on the (010) surface. Surface states exist
outside the projected nodal loop and inside the ellipse.
(c) Enlargement of the double drumhead-like (010) surface bands near
$\tilde{Z}$. (d)-(e) Schematic illustration of the
formation of ``double-drumhead'' surface states due to nodal loop
twisting and intersecting.}
\end{figure}

\paragraph*{Surface states.}
One of the most important signatures of a topological nodal-line
semimetal is the existence of ``drumhead-like'' surface states
either inside or outside the projected nodal loop \cite{TNS,MTC,Cu3PdN,
*Cu3NNi,PbTaSe2,*TlTaSe2}. Figure~\ref{fig4_surf} shows the calculated
local density of states
for the semi-infinite (001) surface, from which we have an intuitive
visualization of surface band structure and Fermi surface.
For the Sn-terminated (001) surface, the drumhead-like surface states
are nestled inside the projected nodal loop as shown in Fig.~\ref{fig4_surf}(a)-(b).
Unlike the nearly flat ``drumhead'' surface states in some previous
works\cite{MTC, Cu3PdN}, the surface states of BaSn$_2$ are more
dispersive, with a bandwidth of about 180\,meV. It is interesting to note
that when the (001) surface is terminated on the Ba atomic layer,
the topological surface states fill the region outside the projected
nodal loop as shown in Fig.~\ref{fig4_surf}(c)-(d), exactly the
opposite as for the Sn-terminated surface.
Such an interesting phenomenon stems from the  properties of 1D atomic
chains with $\mathcal{P}\mathcal{T}$ symmetry, as explained
in the supplement 
\footnotemark[\value{footnote}].
We have also checked that the local density of states of the surface states
decays rapidly into the bulk and becomes negligible within five principal
layers \footnotemark[\value{footnote}].

The surface states for the (010) surface are dramatically different.
The interior region of the projected nodal loop now is divided into
three subregions, one ellipse and two crescents [see Fig.~\ref{fig1_stru}(d)
and the white dashed line in Fig.~\ref{fig5_surf}(b)]. As is shown in
Fig.~\ref{fig5_surf}, there are two surface bands inside the ellipse,
none inside the crescents, and one topological surface band outside
the projected nodal loop.  The single surface band outside the projected
nodal loop is regarded as the hallmark of a bulk topological nodal-loop
semimetal. On the other hand, the two surface bands inside the ellipse
result from the twist and intersection of the projected nodal
loop along the $\tilde{U}$-$\tilde{Z}$ line, so that the surface states
outside the untwisted nodal loop [Fig.~\ref{fig5_surf}(d)] become
overlaid with each other, leading to two drumhead-like surface states
inside the ellipse [Fig.~\ref{fig5_surf}(e)].

\paragraph*{SOC effect.}
In general, spin-orbit coupling (SOC) can drive the nodal-line semimetal
into different topological phases, such as a topological insulator \cite{CaAgX},
3D Dirac semimetal \cite{Cu3PdN,*Cu3NNi}, or other kinds of nodal-line semimetals
\cite{PbTaSe2,*TlTaSe2}. In graphene it is well known that SOC splits
the BT, leading to a quantum spin Hall insulator \cite{KanePRL2005}.
Similarly, in a 3D system with a nodal loop, SOC generally lifts the
degeneracy on the loop and drives the system into a 3D topological
insulator \cite{CaAgX}.
Some crystalline symmetries, such as $C_4$ rotation or reflection, may
protect the existence of the nodes at isolated \textbf{k} points or in
a mirror plane, even in the presence of SOC, resulting in 3D Dirac
semimetals \cite{Cu3PdN,*Cu3NNi} or nodal-line semimetals protected by
crystalline symmetry \cite{PbTaSe2,*TlTaSe2}.

In the case of BaSn$_2$, our first-principles calculations indicate
that the SOC, which is much smaller than the dispersion of
the nodal loop, completely lifts the degeneracy of the nodal loop,
and a topological-insulator phase with a global energy gap of
13\,meV results. We calculate the $\mathbb{Z}_2$ invariants, finding,
$(\nu _0;\nu _1\nu _2\nu _3)=(1;001)$, and the Dirac surface states,
thus identifying the nontrivial band topology \footnotemark[\value{footnote}].
We further check the SOC-induced gap along the nodal loop locally and
find that the gap is $\sim$50\,meV on the $\Gamma$-$M$-$L$-$A$ plane,
while it increases to a maximum value of $\sim$160\,meV on the $A$-$H$ line.
Note that BaSn$_2$ is an extreme case among the $AX_2$ compounds, where
both Ba and Sn are heavy elements with relatively strong SOC.

The SOC strength may be diminished by doping or substituting lighter
elements such as Si or Ca,
potentially allowing the nodal-line semimetal phase to reappear. Our detailed investigations
are reported in the Supplement \footnotemark[\value{footnote}].
Similar to BaSn$_2$, alkaline-earth
stannides (SrSn$_2$ and CaSn$_2$), germanides (BaGe$_2$, SrGe$_2$ and CaGe$_2$)
and silicides (BaSi$_2$ and SrSi$_2$) all exhibit nodal-line semimetal
behaviors in the absence of SOC. Meanwhile, the SOC-induced splitting
of the nodal line gets smaller from stannides to silicides ($<10$\,meV).
Moreover, various new properties show up in these materials. For example,
in addition to the nodal loop, isolated point nodes appear around the $M$ point
in BaSi$_2$; while unlike BaSn$_2$, there are only isolated nodes
on the $k_z$ axis in CaSi$_2$.
More importantly, a type-II 3D
Dirac-fermion electronic structure, which is robust against SOC,
appears in the valence bands of most akaline-earth stannides and germanides.
This new type of quasiparticle is a counterpart of the recently discovered
type-II Weyl fermions \cite{type2Weyl}, and is expected to exhibit novel
properties distinct from previous Dirac semimetals. The details of this
exotic type-II Dirac semimetal will be discussed in future
work \cite{huangduan}.

\paragraph*{Conclusion.}
In summary, we have theoretically predicted the existence of 3D topological
nodal-line semimetals in alkaline-earth compounds $AX_2$ when SOC is neglected.
The single snake-like nodal loop distinguishes this class of materials from
existing nodal-line semimetals which typically have multiple nodal rings.
The drumhead-like surface states on the (001) surface may appear either inside
or outside the projected nodal loop, depending on surface termination,
while some unconventional double-drumhead-like surface states occur on
the (010) surface.
Including SOC drives the system into a topological-insulator phase
whose $\mathbb{Z}_2$ indices are (1;001).
As the SOC effect is much weaker in
alkaline-earth compounds composed of light elements, the nodal-line
semimetal as and its exotic electronic properties are expected to be
observed in these materials.

\begin{acknowledgments}
We thank Professor Hongbin Zhang and Feng Liu for valuable discussions and fruitful suggestions.
H. H. and W. D. are supported by the Ministry of Science and Technology
of China (Grant Nos. 2011CB606405 and 2011CB921901) and the National
Natural Science Foundation of China (Grant No. 11334006). D.V. is supported by
NSF Grant No. DMR-14-08838. J.L. would like to acknowledge the support
from DMR-14-08838 and DMR-15-06119.
\end{acknowledgments}

\providecommand{\noopsort}[1]{}\providecommand{\singleletter}[1]{#1}%

\end{document}